\documentclass[prl,twocolumn]{revtex4}
\usepackage{graphicx,amssymb}
\usepackage{natbib}

\newcommand{\degc}{\ensuremath{^{\circ}}}
\begin{document}
\title{Surface Transport in Pre--Melted Films with Application to Grain--Boundary Grooving}
\author{Robert~W.~Style and M.~Grae~Worster}
\address{Institute of Theoretical Geophysics, Department of Applied Mathematics and Theoretical Physics, University of Cambridge CB3 9EW Cambridge, England}
\begin{abstract}
We present a new model of surface transport in premelted films that is
applicable to a wide range of materials close to their
melting point. We illustrate its use by applying it to the evolution of
a grain boundary groove in a high vapour pressure material and show that 
Mullins's classical equation describing transport driven by gradients in 
surface curvature is
reproduced asymptotically. The microscopic contact angle at the groove root
is found to be modified over a thin boundary layer, and the apparent contact
angle is determined. 
An explicit transport coefficient is derived that governs the evolution rate
of systems controlled by surface transport through premelted films. The
transport coefficient is found to depend on temperature and diverges as the
bulk melting temperature is approached.
\end{abstract}

\maketitle

It is well established \cite{fren85} that melting in any material is initiated at the free
surface. Thus, a molecularly thin layer of melt liquid can exist
at the surface at temperatures below the bulk transition temperature, depending on the
predominant intermolecular interactions at the solid--vapour interface.
Typically, repulsive van
der Waals forces or electrostatic interactions act to thicken the film, while
the film width is constrained by an attractive pressure arising from the liquid
being held below its freezing point. (For a wide ranging review, see Dash et al.
\cite{dash95}.) Wettlaufer and Worster \cite{wett95} have shown that it is possible to use
lubrication theory to model the flow in the film, and they found that their 
results were in good
agreement with experiment, using bulk parameter values (e.g. dynamic viscosity) in the flow \cite{wett96}. Furthermore, there exists much
evidence, both theoretical \cite{trav00},\cite{kopl95} and experimental
\cite{ravi02} showing that the continuum fluid dynamical approach, combined with
bulk parameters can be used for films thicker than about ten molecular diameters.
Therefore, in systems with sufficiently thick films, we can derive a system of
continuum equations to model the evolution of
solid--vapour surfaces near to their melting point.

In this letter, we derive the equations pertaining to surface melting and
flow in a surface--melted film. We then apply these to the evolution of a
grain--boundary (GB) groove and demonstrate that, in the long time limit, the
original equation derived by Mullins \cite{mull57}
\begin{equation}
y_t+By_{xxxx}=0,
\end{equation}
is recovered asymptotically. Here, $y(x,t)$ is the surface height, $x$ is distance parallel to the surface and $B$ is a
constant transport parameter. We use typical bulk parameters for the water/ice
system at $-1\degc$ to give predictions of grooving rates that are in agreement
with typical observations. We focus on ice because it is experimentally accessible and
because there is much research on it due to its unique importance in geoscience. It
is also known to premelt against air \cite{elba93}, and may be doped to yield
sufficiently thick films to justify the continuum approach \cite{bena04}.

Firstly, using the assumption that the system is in local thermodynamic
equilibrium, we have as a consequence of the Gibbs--Duhem relationship
\cite{wood90} that
\begin{equation}
\label{eq-clap}
p_s- p_l = \rho_s q_m \left(\frac{T_m-T}{T_m}\right),
\end{equation}
where $\rho$, $p$, $T$, $T_m$ and $q_m$ are density, pressure, system temperature,
melting temperature and latent heat of fusion respectively, and subscripts $s$ and $l$ correspond
to the solid and liquid phases. We have also used the assumption 
that the system is isothermal.

\begin{figure}[!h]
\centering
\includegraphics[width=8cm]{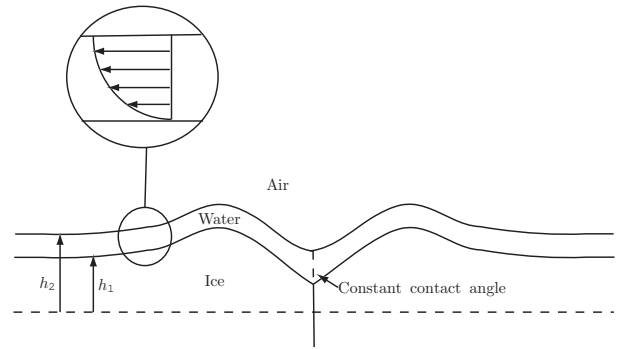}
\caption{Schematic diagram of a grain--boundary groove. The system is isothermal
with $T=T_0$.}
\label{ice}
\end{figure}

Defining $h_1(x,t)$ to be the height of the solid--liquid surface, $h_2(x,t)$ to be
the height of the liquid--air surface, $d(x,t)=h_2-h_1$ to be the thickness of the
liquid layer (cf Fig. \ref{ice} for a specific example), $\gamma_{sl}$ and $\gamma_{la}$ to be the (constant) surface free
energies of the two interfaces, and $A$ to be the effective Hamaker constant
\cite{wett95}, we have that the Helmholtz free energy
\[
F=\int_0^{\infty}\biggl[-p_s(h_2-d)-p_ld+p_ah_2+\gamma_{la}(1+h_2'^2)^{1/2}
\]\begin{equation}
+\left(\gamma_{sl}+\frac{A}{12\pi d^2}\right)(1+h_1'^2)^{1/2}\biggr]dx.
\end{equation}
Treating the integrand as a function of $h_1$ and $d$ and minimising $F$ using the
Euler--Lagrange equation gives
\begin{equation}
p_s-p_l=\frac{A}{6\pi d^3}(1+h_1'^2)^{1/2}+\left(\gamma_{sl}+\frac{A}{12\pi d^2}\right)K_1
\end{equation}
where $K_1$ is the curvature of the solid--liquid interface $z=h_1$.

We assume that the thickness of the film is much greater than a 
characteristic molecular dimension, and that the slopes of the surfaces are
small. Therefore
\begin{equation}
\gamma_{sl}\gg\frac{A}{12\pi d^2},\quad h_1''\ll1,
\end{equation}
and we can approximate
\begin{equation}
p_s-p_l = \frac{A}{6\pi d^3}-\gamma_{sl}h_1'',
\end{equation}
which we combine with Eq. (\ref{eq-clap}) to give a general result for a surface--melted film 
(cf \cite{dash95})
\begin{equation}
\label{eq-en}
\frac{A}{6\pi d_0^3}\equiv\rho_s q_m\left(\frac{T_m-T_0}{T_m}\right)=-\gamma_{sl}h_1''+\frac{A}{6\pi d^3}.
\end{equation}

The thickness of the layer
is determined by the competition between the rise in energy required
to sustain a liquid film below the melting point, and the van der Waals (VdW) forces, as
expressed in the balance between the LHS and last term on the RHS in
(\ref{eq-en}). As can also be seen in the equation, the Gibbs--Thomson effect
manifests itself in that
an increase in surface curvature causes an increase in the thickness of the
film.

The second Euler--Lagrange relationship from above yields
\begin{equation}
\label{eq-al}
p_l=p_a -\gamma_{la}h_2''-\frac{A}{6\pi d^3}.
\end{equation}

So, in regions where the film is thick and has high curvature, the
liquid pressure is high. This drives flow away, towards regions where the
curvature is lower and the film thinner.

Finally, assuming that the thickness of the film is
sufficiently small, we can use lubrication
theory \cite{batc67} to give the velocity in the layer (see Fig.
(\ref{ice}) insert)
using a no--slip condition at the solid--liquid interface and a stress-free
condition at the liquid--air interface.
Mass conservation is expressed by
\begin{equation}
\label{eq-mass}
\frac{\partial q}{\partial x} + \dot{h}_2=0,
\end{equation}
where $q(x)=-d^3 p_x/(3\mu)$, and $\mu$ is the dynamic viscosity of water.

We introduce dimensionless variables with  a lengthscale determined
by taking the balance between VdW and curvature terms in equation (\ref{eq-en})
to give

\begin{equation}x=\left(\frac{A}{6\pi\gamma_{sl}}\right)^{1/2}\xi\equiv\epsilon\xi\end{equation}
\begin{equation}t=\left(\frac{3\mu^2A}{2\pi\gamma_{sl}^3}\right)^{1/2}\tau=\delta\tau\end{equation}
and we write $(h_1, h_2, d, d_0)= \epsilon({\cal H}_1, {\cal H}, {\cal D}, {\cal D}_0)$.

Thus we obtain a coupled pair of dimensionless partial differential equations
\begin{equation}
\label{eq1}
\frac{1}{{\cal D}_0^3}-\frac{1}{{\cal D}^3}=-{\cal H}''+{\cal D}'',
\end{equation}
\begin{equation}
\label{eq2}
\left[{\cal D}^3\left(\gamma {\cal H}'''+\left(\frac{1}{{\cal D}^3}\right)'\right)\right]'+
\dot{{\cal H}}=0,
\end{equation}
where $\gamma=\gamma_{la}/\gamma_{sl}$. Boundary conditions for these equations
depend on the specific problem to be addressed.

We can now, for example, address the problem of a grain--boundary groove. The model to be
analysed is shown in Fig. \ref{ice}.  Two ice crystals, symmetric about the GB between them are exposed to the air. The system is maintained
at a uniform temperature
$T_0$ sufficiently close to the melting point $T_m$ that surface melting occurs
and a liquid film is formed, separating solid from air. Initially, the upper
surfaces of both
crystals are planar, but equilibrium demands a fixed dihedral angle
(given by the Young-Dupr$\acute{\mathrm{e}}$ relationship \cite{land58}),
at the tri-junction formed by the
GB and the solid--liquid interfaces. The premelted film is thereby made thicker
near the GB, the disjoining force across it is correspondingly weaker and the
liquid pressure higher. This
drives a flow in the premelted film transporting mass outwards to form a GB
groove.

Using this model, we take boundary conditions to equations (\ref{eq1}),(\ref{eq2})
to be ${\cal D}(\infty)={\cal D}_0$,
${\cal H}'(0)=0$, ${\cal D}'(0)=-\alpha$ and $q(0)=0$, corresponding
respectively to constant film thickness at infinity, constant
contact angle at the groove base (with cotangent $\alpha$), continuous derivative of the liquid--air
interface above the groove and zero
mass flux across the plane of symmetry.

As time proceeds, the groove widens and deepens, while the thickness of the premelted film stays
relatively constant. Therefore we can make the assumption (justifiable \emph{a
posteriori}) that ${\cal D}\ll {\cal H}$ and hence ${\cal D}''\ll {\cal H}''$
away from a neighbourhood of the $z$ axis. Thus (\ref{eq1}) becomes
\begin{equation}
\frac{1}{{\cal D}_0^3}-\frac{1}{{\cal D}^3}=-{\cal H}''.
\end{equation}
Therefore, assuming that ${\cal D}\approx {\cal D}_0$ (see below) and combining with (\ref{eq2}), we recover the
Mullins equation
\begin{equation}
\label{mulli}
{\cal H}_{\tau}+{\cal D}_0^3(\gamma+1){\cal H}_{\xi\xi\xi\xi}=0,
\end{equation}
with dimensional transport coefficient
\begin{equation}
\label{transport}
B\equiv\frac{A(\gamma+1)\gamma_{sl}}{18\pi\mu\rho_sq_m}\frac{T_m}{(T_m-T_0)}.
\end{equation}

This equation admits a similarity solution \cite{mull57}
\begin{equation}
\label{ss}
{\cal H}_{s}=[{\cal D}_0^3(\gamma+1)\tau]^{1/4}f(\eta)
\end{equation}
where
\begin{equation}
\eta=\frac{\xi}{[{\cal D}_0^3(\gamma+1)\tau]^{1/4}}.\end{equation}
and the function $f$ satisfies
\begin{equation}
\label{simi}
f^{\mathrm{iv}}+\frac{f}{4}-\frac{\eta f'}{4}=0,
\end{equation}
subject to the boundary conditions $f'''(0)=0$, $f'(0)=\beta$, $f'(\infty)=0$
and $f(\infty)=0$. This yields ${\cal H}_s(0)$ which we will use to match to
the inner boundary layer solution. The constant $\beta$ is the cotangent of the apparent contact angle seen by this outer region, and will be determined by
matching to the inner asymptotic solution, derived as follows.

In the similarity solution regime, we take ${\cal H}\sim\tau^{1/4}f(\eta)$ and
${\cal D}\sim g(\eta)$. Combining these estimates with Eq. (\ref{eq1}) yields
${\cal D}={\cal D}_0+O(\tau^{1/4})$ and hence the similarity solution is expected to break
down at the GB when ${\cal D}'\rightarrow-\alpha$.

In this regime, we assume that the $\tau^{1/4}$ time dependence of the similarity solution is carried over
into the inner solution. Then consideration of the order of magnitude of
terms in Eq. (\ref{eq2}) shows that ${\cal H}_{\tau}$ vanishes for large times, and we can make
quasi-steady approximations of ${\cal H}(\xi,\tau)=C\tau^{1/4}+H(\xi)$ and
${\cal D}(\xi,\tau)=D(\xi)$ inside the boundary layer.

Applying a zero flux boundary condition at the GB, Eq. (\ref{eq2}) becomes
\begin{equation}
D^3\left[\gamma H'''+\left(\frac{1}{D^3}\right)'\right]=0,
\end{equation}
which we integrate with $H''(\infty)=D''(\infty)=0$, $D'(0)=-\alpha$ and
$W'(\infty)=0$ to obtain
\begin{equation}
\label{inner}
{\cal H}_{inner}=C\tau^{1/4}+\frac{D}{1+\gamma}+\frac{\alpha\xi}{1+\gamma}+c,
\end{equation}
where, from Eq. (\ref{eq1}), $D$ satisfies the equation
\begin{equation}
\label{fluxcond}
(1+\gamma)D''=\gamma\left(\frac{1}{{\cal D}_0^3}-\frac{1}{D^3}\right).
\end{equation}

Substituting ${\cal H}_s=\tau^{1/4}f(\eta)$ into Eq. (\ref{eq2}), we find that as
$\tau\rightarrow\infty$, ${\cal H}_s$ satisfies the full equation. Therefore we can match Eqs.
(\ref{ss}) and (\ref{inner}) by letting ${\cal H}_{inner}(\xi)\rightarrow {\cal H}_s$ and
$D(\xi)\rightarrow {\cal D}_0$ for large $\xi$ (the outer edge of the inner solution),
and matching this to the small $\xi$ limit of Eq. (\ref{mulli}).

For small $\xi$, Eq. (\ref{ss}) becomes
\begin{equation}
\label{ss2}
{\cal H}_s={\cal H}_s(0)+\xi\beta+O(\tau^{-1/4}),
\end{equation}
giving that
\begin{equation}
\label{match}
{\cal H}_{inner}={\cal H}_s(0)+\frac{D-{\cal D}_0}{1+\gamma}+\frac{\alpha\xi}{1+\gamma}.
\end{equation}
Importantly, we have a value for the apparent cotangent
\begin{equation}
\label{beta}
\beta=\frac{\alpha}{1+\gamma}.
\end{equation}
Eq. (\ref{beta}) shows that for $\gamma_{la}$ large relative
to $\gamma_{sl}$, the system will try to minimize the liquid--air
interfacial area, and hence $\beta\rightarrow0$. Conversely for $\gamma_{sl}\gg\gamma_{la}$,
$\beta\rightarrow\alpha$ as the solid--liquid interfacial curvature is minimized
at the GB to reduce surface area.
This also implies that the apparent cotangent $\beta$ can differ
substantially from the actual cotangent $\alpha$ suggesting
that experimental determination of $\alpha$ could yield incorrect results
due to the apparent contact angle being measured instead.

\begin{table}
\caption{Table of typical values for ice/water/air at $T=-1\degc$}
\label{tov}
\begin{center}
\begin{tabular}{|l|r|r|}
\hline
Constant & Value & Units \\ \hline
$\rho_s$ & $917$ & kg\,$\mathrm{m}^{-3}$ \\
$q_m$ & $3.34\times10^{5}$ & $\mathrm{J}\,\mathrm{kg}^{-1}$ \\
$\gamma_{sl}$ & $3\times10^{-2}$ & J\,$\mathrm{m}^{-2}$ \\
$\gamma_{la}$ & $7.57\times10^{-2}$ & J\,$\mathrm{m}^{-2}$ \\
$A$ & $1.50\times10^{-20}$ & J \\
$\epsilon$ & $5.16\times10^{-10}$ & $\mathrm{m}$ \\
$\delta$ & $2.9\times10^{-11}$ & $\mathrm{s}$ \\
$d_0$ & $8.93\times10^{-10}$ & m \\
$w_0$ & $1.73$ & - \\
$\mu$ & $1.787\times10^{-3}$ & $\mathrm{kg}\,\mathrm{m}^{-1}\,\mathrm{s}^{-1}$ \\
$B$ & $1.40\times10^{-26}$ & $\mathrm{m}^4\,\mathrm{s}^{-1}$\\
\hline
\end{tabular}
\end{center}
\end{table}

To solve the complete system of equations, we used an explicit numerical scheme. At each
step, the ${\cal D}$ profile is calculated from (\ref{eq1}) by a relaxation
technique. The new value of ${\cal H}$ at $(\tau+d\tau)$ can then be calculated from (\ref{eq2}),
using the previous values of ${\cal H}(\tau)$ and ${\cal D}(\tau)$.
\begin{figure}[!h]
\centering
\includegraphics[width=8cm]{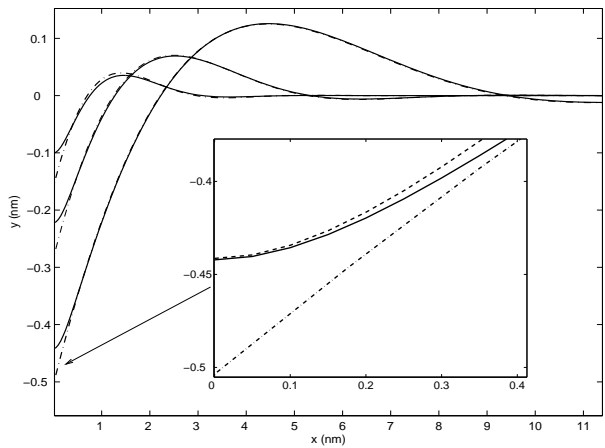}
\caption{Numerical solutions (continuous lines) and similarity solutions (dashed lines) for time evolution of a groove at a liquid--air interface. $y$ is the the height of the interfaces relative to the initial position. Graphs are plotted in terms of $x$ and $t$
at $t=8.7\times10^{-7},1.7\times10^{-5},2.9\times10^{-4}$. Inset shows numerical solution (continuous line), similarity solution (dashed line) and inner boundary layer asymptotics (dash-dot line) for the inner
boundary layer regime at $t=2.9\times10^{-4}$.
Parameters were chosen as ${\cal D}_0=1$, $\gamma=0.5$.}
\label{mull}
\end{figure}

Fig. \ref{mull} shows the time evolution of the groove. Good agreement can
be seen, even at small times, between the numerical solutions (continuous lines)
and the similarity solution (dashed lines).
Boundary
conditions for the similarity solution are ${\cal H}'(0)=\beta$, $q(0)=0$,
${\cal H}(\infty)=0$ and ${\cal H}'(\infty)=0$.
The inset figure shows the inner boundary layer region for $t=2.9\times10^{-4}$.
The inner asymptotics (dash--dot) are shown, and also show good agreement with
the numerical solution.

From Eq.(\ref{transport}) we see that the rate--controlling coefficient is 
inversely proportional to
the reduced temperature, and grows as $T_0$ approaches the melting point.
We also see that less viscous, less dense liquids with
thicker layers ($A$ large) and stronger surface tension cause faster grooving
than more viscous, dense liquids with thin layers and weak surface tension.
Typical parameters for ice at $-1\degc$ (see Table \ref{tov}) give $B$ to be
approximately $1.4\times10^{-26}\mathrm{m}^4\mathrm{s}^{-1}$ and so typical lengthscales (proportional to
$(Bt^{1/4})$) are of order $0.35\mu$m in
$1$s and $3.5\mu$m in $3$ hours, which is in line with observations of ice \cite{barn02}. At this time, the authors are unaware of any experimental measurements
of $B$ in any material within a suitable temperature range.

The analysis presented above is relevant to any material that premelts against
its atmosphere and forms a constant contact angle at the groove root, in circumstances where it
is maintained at a suitable temperature that premelting can occur. However,
several points should be noted.
(i) We have assumed that the slope of the interfaces is always small, and 
therefore some detail of the form of the groove root may
have been lost. (ii) We assumed that there is a complete phase change
across the surface $z={\cal H}_1$, while the transition may actually occur across
several atomic diameters and involve some level of ordering(eg \cite{huis97}). However, provided the premelted
film is thick enough, the affects of this should be negligible and we may use
continuum approximations. Appropriately thick films may be ensured by the
addition of dopants to the system \cite{bena04}. (iii) We have considered
surface melting controlled by unretarded VdW forces. This can be
easily generalised to retarded VdW forces and electrostatic interactions \cite{wett96}. (iv) We require $\alpha$ to be small for the
lubrication approximation to be valid in the inner region. However, groove root
angles for ice may be very small, so this analysis can break down near the
groove root.

So, in conclusion, we have demonstrated a new mechanism for mass transport on
the surface of a crystal in a suitably high temperature regime. Surface melting
is a ubiquitous phenomena, and so the equations that we have derived should be
applicable to a wide range of materials and situations. We have illustrated the
mechanism of surface transport by applying it to the case of a grain--boundary groove and
demonstrated that it yields Mullins's equation in the asymptotic limit,
although the dihedral angle determined by the surface energies is modified
over the boundary layer by a factor $1/(1+\gamma)$. We note that much richer
behaviour will result from effects such as crystal anisotropy,
impurity content, asymmetry and other grooving mechanisms \cite{mull57}.

The most important result of this letter is that we have found an explicit transport
parameter $B$ (Eq. (\ref{transport})) that controls the rate of surface transport in the
system. It should be noted that $B$ does not depend on the geometry of the grain--boundary
groove. Therefore this parameter should be relevant to most problems involving 
surface melting such as sintering, grain--boundary groove migration, crystal growth and
annealing to name but a few.

The authors would like to thank J.S.Wettlaufer for his critical readings of the
letter.

\end{document}